\documentclass[journal]{IEEEtran}
\usepackage[T1]{fontenc}
\usepackage{color}
\usepackage{flushend}
\usepackage{amsmath}
\usepackage{cite}
\usepackage{algpseudocode} 

\usepackage{booktabs}
\usepackage{tabu}
\usepackage[cmintegrals]{newtxmath}
\usepackage{bm}
\usepackage{indentfirst}
         
\usepackage{amsmath,amssymb}
\usepackage{array}
\usepackage{mathrsfs}
\usepackage{cite}
\usepackage{tikz}
\usepackage{color}
\usepackage{amssymb}
\usepackage{url}
\usepackage{hyperref}
\usepackage{todonotes}
\usepackage{color}
\usepackage{bbding}
\usepackage{multirow}
\usepackage{colortbl}
\usepackage{amsmath}
\usepackage[ruled]{algorithm2e} 
\usepackage{graphicx}
\usepackage{subcaption}

\begin{document}

\title{Image Provenance Analysis via\\Graph Encoding with Vision Transformer}

\author{Keyang~Zhang, Chenqi~Kong,~\IEEEmembership{Member,~IEEE},
Shiqi~Wang,~\IEEEmembership{Senior Member,~IEEE},
Anderson~Rocha,~\IEEEmembership{Fellow,~IEEE}, Haoliang~Li,~\IEEEmembership{Member,~IEEE}
\thanks{K. Zhang is with the Department of Electrical Engineering, City University of Hong Kong, Hong Kong SAR. (email: keyazhang4-c@my.cityu.edu.hk).}
\thanks{C. Kong is with the Rapid-Rich Object Search (ROSE) Lab, School of Electrical and Electronic Engineering, Nanyang Technology University, Singapore. (email: chenqi.kong@ntu.edu.sg.)}
\thanks{S. Wang is with the Department of Computer Science, City University of Hong Kong, Hong Kong SAR. (email: shiqiwang@cityu.edu.hk).}
\thanks{A. Rocha is with the Artificial Intelligence Lab. (\texttt{Recod.ai}) at the University of Campinas, Campinas 13084-851, Brazil (e-mail: arrocha@unicamp.br), URL: \url{http://recod.ai}}
\thanks{H. Li is with the Department of Electrical Engineering, City University of Hong Kong, Hong Kong SAR. (email: haoliang.li@cityu.edu.hk).}}

\markboth{Submitted to IEEE Transactions on Information Forensics and Security}%
{Shell \MakeLowercase{\textit{et al.}}: Bare Demo of IEEEtran.cls for IEEE Communications Society Journals}

\maketitle

\begin{abstract}
Recent advances in AI-powered image editing tools have significantly lowered the barrier to image modification, raising pressing security concerns those related to spreading misinformation and disinformation on social platforms. Image provenance analysis is crucial in this context, as it identifies relevant images within a database and constructs a relationship graph by mining hidden manipulation and transformation cues, thereby providing concrete evidence chains. This paper introduces a novel end-to-end deep learning framework designed to explore the structural information of provenance graphs. Our proposed method distinguishes from previous approaches in two main ways. First, unlike earlier methods that rely on prior knowledge and have limited generalizability, our framework relies upon a patch attention mechanism to capture image provenance clues for local manipulations and global transformations, thereby enhancing graph construction performance. Second, while previous methods primarily focus on identifying tampering traces only between image pairs, they often overlook the hidden information embedded in the topology of the provenance graph. Our approach aligns the model training objectives with the final graph construction task, incorporating the overall structural information of the graph into the training process. We integrate graph structure information with the attention mechanism, enabling precise determination of the direction of transformation. Experimental results show the superiority of the proposed method over previous approaches, underscoring its effectiveness in addressing the challenges of image provenance analysis.
\end{abstract}


\IEEEpeerreviewmaketitle

\section{Introduction}

\IEEEPARstart {I}{n} today's platforms of social media, visual content has emerged as a predominant mode of communication \cite{adami2016social}, transcending linguistic barriers and enabling instantaneous sharing of life experiences, news, and ideas. However, this convenience comes with a hidden risk: maliciously tampered photos can significantly mislead public opinion and contribute to spreading fake news \cite{cao2020exploring, kong2022digital}, posing multiple potential dangers. Extensive research in the prior art has focused on image authentication \cite{han2010content} and image forgery localization \cite{piva2013overview,kong2022detect,guillaro2023trufor,kong2023pixel}. However, with the rapid proliferation of image content circulating on social media platforms, it is of paramount importance to further identify image provenance and construct a provenance graph to ensure image integrity.

\begin{figure}[t]
\centering
\includegraphics[width=\linewidth]{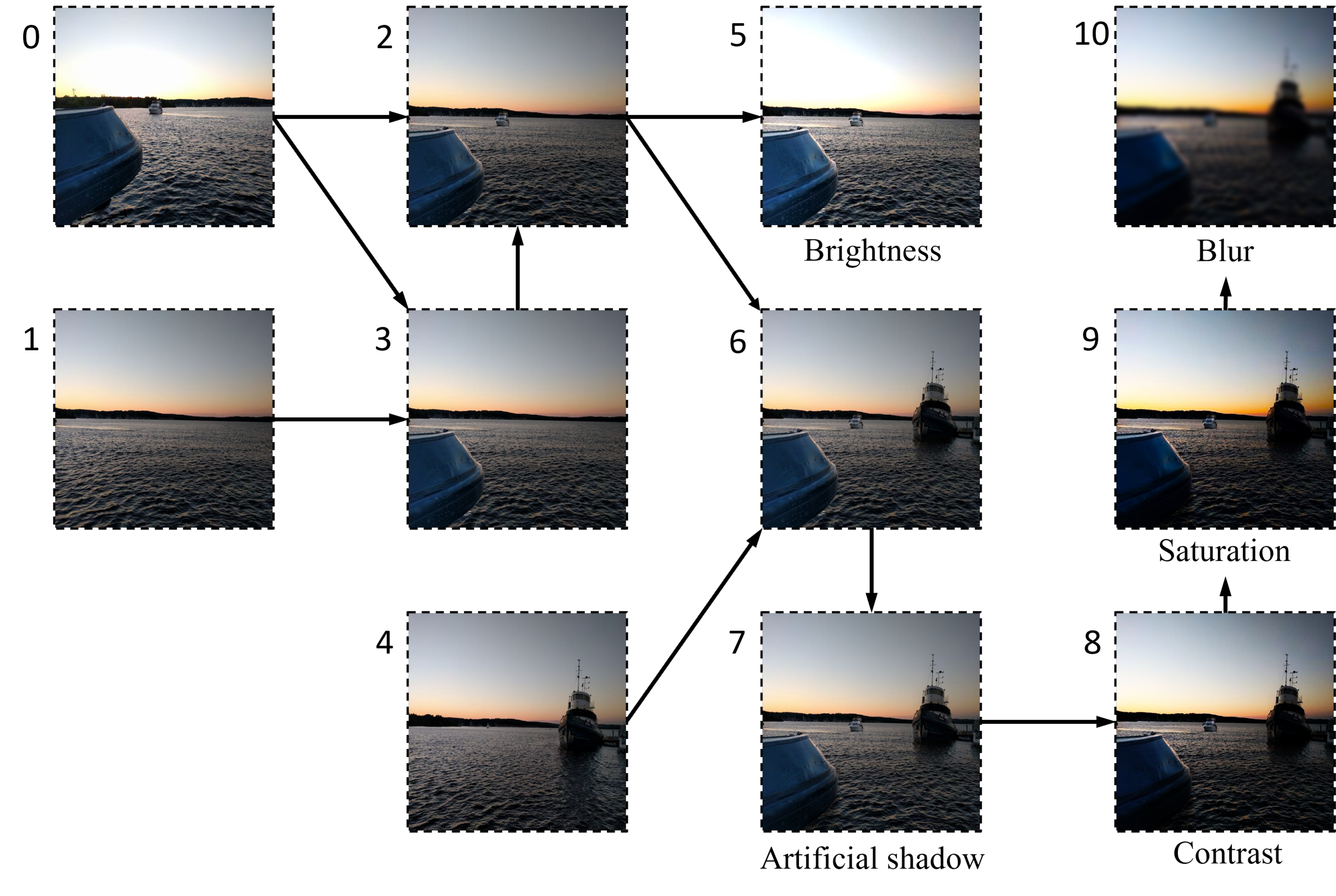}
\caption{Example of a provenance graph, where images are represented as nodes and manipulation relationships are denoted by arrowed edges. The graph originates from base images, which serve as donors of backgrounds or objects to the offspring images. Labels under the images indicate that these are the resultant images from specific types of global transformations. The unlabeled images are generated by compositing content from the donor images.
}
\label{Provenance}
\end{figure}

In response to these challenges, the field of Image Provenance Analysis \cite{moreira2018image} has recently emerged as multimedia phylogeny with focus on visualizing the underlying relationships among related images as illustrated in Fig.~\ref{Provenance}. Provenance analysis goes beyond the simple binary classification of images as fake or real \cite{robertson2019manipulation}. It seeks to reveal the hidden narratives of a set of semantically-similar images, which can more powerfully indicate manipulated intent and provide a comprehensive visual graph. Enhancing our understanding of the context and lineage of digital images strengthens the tools available for maintaining the integrity of visual information, protecting copyright by giving a more concrete evidence chain, and accurately reducing tampered images uploaded to the social networking \cite{liu2014digital} with malicious purposes, which is indistinguishable for forensics tools.

Image provenance analysis aims to identify images within a database that may share manipulation relationships with each other \cite{bharati2019beyond}. The process involves unveiling the relationships among associated images using either directed \cite{moreira2018image,bharati2019beyond,zhang2020discovering} or undirected links \cite{bharati2021transformation} and visually representing these relationships through a provenance graph as illustrated in Fig.~\ref{Provenance}. The procedure for conducting end-to-end provenance analysis is typically divided into two main stages: image filtering/selection and graph construction. As existing image filtering methods \cite{moreira2018image,bharati2019beyond,zhang2020discovering} have achieved outstanding performance in selecting semantically-similar images from databases, this study focuses on the second stage, graph construction, which leverages pre-identified related images to map out the network of image manipulation.

A connected provenance graph is constructed by depicting significant transformation relationships between similar images as links and determining their directions. These relationships encompass local transformations such as splice-paste \cite{huh2018fighting} and global transformations such as blur, noise, or contrast changes \cite{797584}. Link prediction methods primarily rely on similarity calculations between image pairs, using features ranging from traditional descriptors like Scale-Invariant Feature Transform (SIFT) \cite{lowe2004distinctive} and Speeded-Up Robust Features (SURF) \cite{bay2006surf} to more advanced Convolutional Neural Network (CNN) learned features \cite{he2016deep,bharati2021transformation}. The selection of features significantly affects the ability to capture different image transformations and relationships. Link direction determination, a distinct step following link prediction, varies based on image attributes. While some methods analyze metadata \cite{bharati2019beyond}, more robust visual content-based approaches \cite{moreira2018image, zhang2020discovering} focus on inherent image information. These methods either calculate mutual information of pixel values \cite{moreira2018image} or detect forgery traces \cite{zhang2020discovering}, reducing reliance on volatile metadata. Constructing a directed provenance graph allows for tracing content within complex image networks, providing valuable insights into the intent behind digital manipulations.

Nonetheless, existing provenance analysis methods still face challenges during the graph construction stage, which can hinder the effectiveness and accuracy of the analysis:
\begin{itemize}
\item \textbf{Underutilization of graph structure:} Most current methods focus primarily on isolated image content or relationships between individual image pairs, often neglecting the more informative structure of the provenance graph, resulting in limited graph construction performance.
\item \textbf{Limited scope of manipulation and transformation detection:} Existing visual content-based methods tend to focus narrowly on detecting either local manipulations (forgery traces) or global transformations (large-scale modifications), rarely addressing both in tomdein.
\item \textbf{Susceptible link direction interference:} Metadata-Ebased methods are vulnerable to simple modifications or information loss during image format conversion. Existing visual-based approaches heavily depend on pixel histograms or forensic integrity, which often fail to accurately predict link directions between source and target images within the constructed graph.
\end{itemize}

This paper introduces a novel image provenance analysis model to address the abovementioned challenges. Our approach enables end-to-end inference of the entire directed provenance graph, out-performing previous methods focusing solely on either link prediction or direction determination. The key contributions of our work are threefold:

\begin{itemize}
\item We propose an innovative image provenance analysis framework that simultaneously processes all images and integrates graph topology into transformer architectures. This effectively captures the rich information within the entire graph.
\item We establish a new paradigm in link prediction through a designed weighted patch distance learning module, coupled with pretrained model-guided patch weights and whole graph path-length loss, effectively capturing local and global manipulation traces.
\item We design a link direction determination approach by introducing learnable precedence embeddings, graph-structured attention masks, and auxiliary virtual nodes, establishing a new benchmark in predicting directional flow within complex provenance graphs.
\item Quantitative and qualitative results show that our method outperforms existing approaches in accuracy, generalization, and robustness across diverse provenance scenarios, underscoring its effectiveness.
\end{itemize}

In this paper, Sec. II comprehensively reviews previous literature on image provenance analysis and graph construction. Sec. III details the proposed provenance graph construction framework. Sec. IV elaborates on the experimental setups, data augmentation techniques, and the baseline models employed. Sec. V presents extensive experimental results compared with other baselines. Finally, Sec. VI concludes the paper and discusses limitations and possible future research directions.

\section{Related Work}
\subsection{Image Manipulation Detection}
To combat the threats of the spread of disinformation through manipulated images, various manipulation detection techniques have been developed \cite{thakur2020recent}. The most widely researched field is image forgery detection, which aims differentiating manipulated images from authentic ones \cite{rossler2019faceforensics++} and precisely localize forgery areas within an image \cite{guillaro2023trufor}. Forgery detection primarily focuses on revealing alterations to the content of images \cite{mehrjardi2023survey}, such as copy-move \cite{chauhan2016survey}, splicing \cite{zhao2011detecting}, and inpainting \cite{wu2021iid}. In turn, image transformations do not alter the content directly, such as variation, in contrast, \cite{shan2019robust}, resampling \cite{bunk2017detection} and JPEG compression \cite{amerini2017localization}, which are often regarded to conceal traces of forgery, thereby suggesting potential tampering.

Lately, there has been a significant shift in manipulation detection methods, veering away from traditional forensic techniques that rely on manual features among blocks or key points. Instead, the field has embraced more robust learning-based approaches that can identify a broader range of manipulations. These methods leverage high-level feature extraction to analyze inconsistencies in invisible fingerprints across the image and detect traces of digital manipulations. Such techniques demonstrate enhanced effectiveness in identifying and addressing various digital image alterations. Recent research highlights that more than merely distinguishing fake images is required; understanding the sequential history of multi-step operations is crucial in certain scenarios. However, previous forensics methods focusing on single-image-based manipulation detection fall short for social media platforms. These approaches need to clarify the intent behind manipulations, which is essential for countering the dissemination of misinformation and disinformation.

\subsection{Provenance Analysis}
Provenance Analysis was first introduced in the context of Internet image archaeology by Kennedy \textit{et al.} \cite{kennedy2008internet}, which aimed to construct relationships among related Internet-published images through identifiable manipulation features and pairwise comparison. Subsequently, Dias \textit{et al.} \cite{dias2011image} proposed the earliest general process for provenance analysis that includes building a dissimilarity matrix and employing a spanning tree algorithm. However, this study focused exclusively on single-root provenance graphs constructed from different versions of a single base image and considered only a limited range of transformations. The Open Media Forensics Challenge \cite{guan2019mfc} has been established to provide large-scale provenance datasets featuring various manipulation techniques and graph topologies. Leveraging these resources, Bharati et al. \cite{bharati2017u,moreira2018image} integrate content-based image retrieval \cite{pinto2017provenance} with dissimilarity computations, which take into account the number of matching interest points. Furthermore, recent research \cite{bharati2019beyond,moreira2018image,zhang2020discovering} underscores the importance of the direction of transformations in accurately interpreting the viewpoints behind manipulations. Efforts include using asymmetric mutual information \cite{moreira2018image} derived from the pixel values of image pairs to delineate directional relationships. Despite these advancements, earlier methodologies relying on handcrafted features faced limitations, including vulnerability to various types of manipulation and reduced robustness against noise. To address these challenges, a metadata-based method \cite{bharati2019beyond} has been developed. This approach harnesses external context information, enhancing computational efficiency and accuracy. However, its effectiveness is contingent upon the integrity of metadata and faces tampering threats in practical applications.

Recent advances in deep learning have significantly improved the performance of image provenance analysis by leveraging learnable image features instead of handcrafted features. A notable development for undirected link prediction is introducing a ranking-based framework \cite{bharati2021transformation}, which is trained on custom-designed sets of quadruplet images and utilizes image embeddings to quantify the number of transformation steps. Building on the capabilities of image forgery detection systems and large-scale forensics datasets, Zhang \textit{et al.} \cite{zhang2020discovering} have developed a hybrid network. This network simultaneously evaluates the integrity of images \cite{sabir2018deep} and assesses local changes between images, which helps determine the direction of transformations. Despite these innovations, the method presents challenges, including risks of mismatched manipulation types and limited scalability. Furthermore, there is a noticeable gap in the current landscape of deep learning applications in image provenance analysis: the absence of a comprehensive deep learning framework that integrates both link prediction and the identification of source and target images.

\subsection{Graph Construction}
Provenance Analysis extends beyond simple visual content processing and relies on graph construction techniques. These techniques are crucial for establishing and visualizing the provenance graph. Graph construction provides a systematic way to analyze data structured in graph form, applicable in diverse domains like social networks, point clouds of shapes \cite{wang2019dynamic}, and chemoinformatics \cite{duvenaud2015convolutional}. This process involves predicting isolated links or modeling the overall graph topology \cite{qiao2018data}, which is vital for uncovering potential relationships within the data. Traditional methods focus on computing similarities based on explicit features to construct various graphs, including spanning trees \cite{stam2014modern} and relative neighborhood graphs \cite{toussaint1980relative}. However, with advancements in deep learning, graph neural networks (GNNs) such as RecGNN \cite{ruiz2020gated}, ConvGNN \cite{scarselli2008graph}, and Graph Attention Network (GAT) \cite{velivckovic2017graph} have emerged. These networks leverage graphs' inherent topology and features to infer relationships within the data more effectively. Recent studies \cite{mialon2021graphit, dwivedi2020generalization} have extended the capabilities of the Transformer network to handle arbitrary graph data, demonstrating competitive performance. This evolution means a significant shift towards more dynamic and sophisticated methods in graph-based data analysis within image provenance tasks.

\section{Proposed Method}

\subsection{Preliminaries}
The objective of provenance analysis is to create a provenance graph $G(V,E)$, where the vertices $V$ represent images included in each provenance case under analysis, and the edges $E$ illustrate the transformational relationships between these images. The process begins with selecting images from a large-scale database that share content with the probe image or its content donors. In our task, the initial filtering is performed in advance, and all images related to the probe are provided. Subsequently, the primary transformation relationships are identified and represented as edges, assembled into an undirected provenance graph. This graph is further refined into a directed format by assigning directions to each edge, indicating the source and target of each transformation. The resulting directed graph provides a comprehensive representation of the image manipulation history, offering insights into the sequence of manipulations and potentially revealing underlying intentions and messaging.

\begin{figure}[t]
\centering
\includegraphics[width=\linewidth]{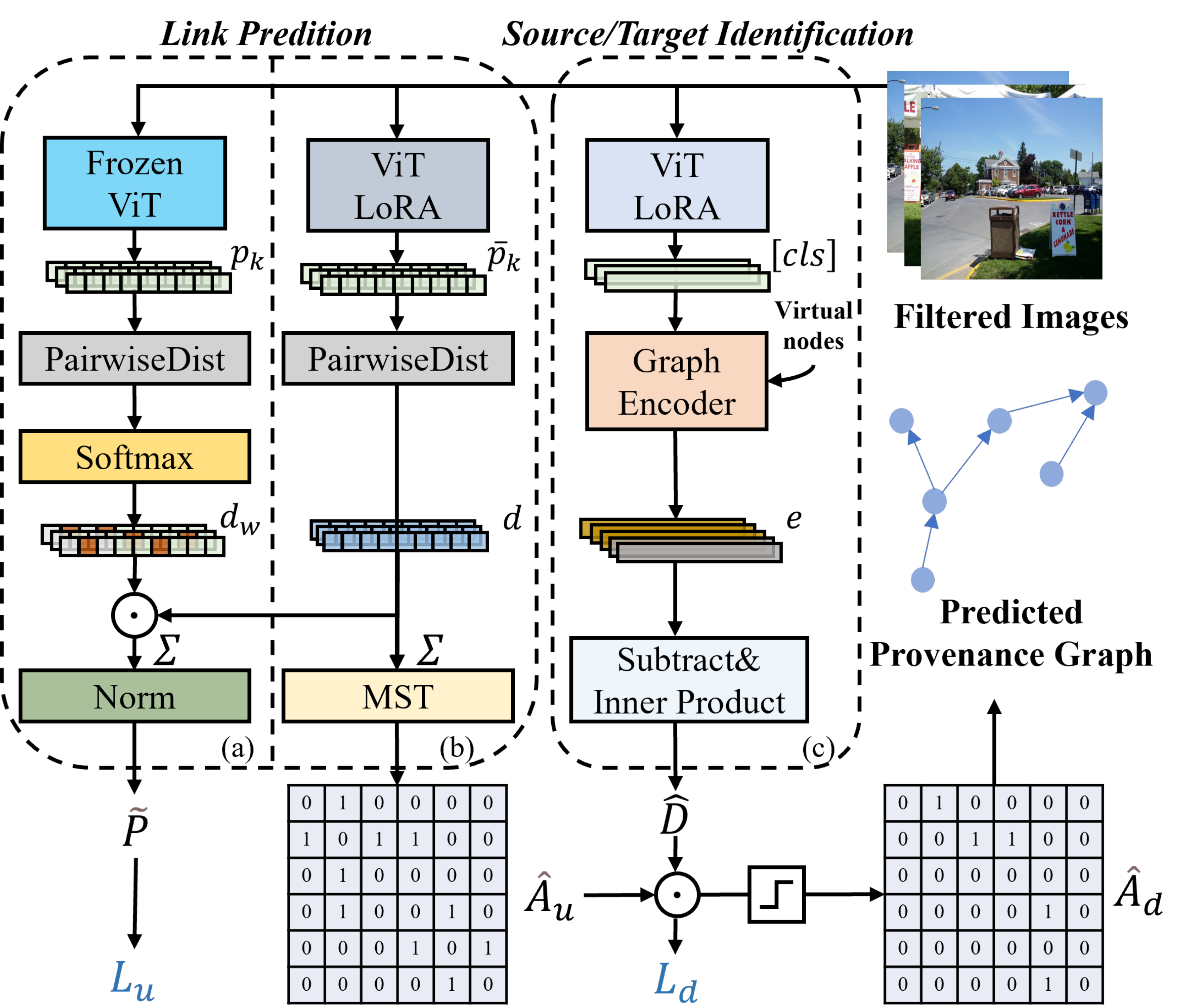}
\caption{Overview of proposed provenance analysis framework to construct the directed provenance graph from a set of semantically-similar images. (a) Determine the weights of patches via softmax applied to pairwise distances of pre-trained ViT patch embeddings. These weights emphasize regions manipulated in the images, influencing the loss function.
(b) Generate learnable ViT-LoRA patch embeddings for computing dissimilarities between images and inferring the undirected adjacency matrix \(\hat{A}_u\) using the Minimum Spanning Tree (MST) algorithm.
(c) Construct the direction matrix \(\hat{D}\) by analyzing image precedence and virtual node embeddings from the graph encoder. The final directed provenance graph, represented as \(\hat{A}_d\), is predicted by fusing \(\hat{A}_u\) and \(\hat{D}\).}
\label{Ovewview}
\end{figure}

\subsection{Overall framework}
As illustrated in Fig.~\ref{Ovewview}, our method for provenance analysis is designed as an end-to-end system that generates a directed provenance graph from a set of images. The designed framework incorporates an undirected link prediction module and a link direction determination module. For the undirected link prediction, depicted in Fig.~\ref{Ovewview} (a), we begin by extracting patch embeddings using a Vision Transformer (ViT) \cite{dosovitskiy2020image} pre-trained on ImageNet, and apply the Low-rank Adaptation (LoRA) \cite{hu2021lora} layer for fine-tuning. The distance between embeddings reflects the similarities of images and is targeted to fit the actual graph structure. During the training phase, as shown in Fig.~\ref{Ovewview} (b), patch weights from the pre-trained ViT model are applied to help localize the manipulated regions. Moving to the source/target identification phase, shown in Fig.~\ref{Ovewview} (c), image embeddings generated by a second feature extractor are processed by the Graph Structure Masked Attention Encoder. It produces the precedence embeddings for each image to represent its provenance position in the latent space and the virtual embeddings to help construct the direction matrix $\hat{D}$. The final step involves element-wise multiplication of $\hat{A}_u$ and $\hat{D}$ to form the directed adjacency matrix $\hat{A}_d$, which represents the directed provenance graph.

\subsection{Undirected Link Prediction}
Constructing a provenance graph fundamentally relies on identifying the manipulation relationships. In the link prediction stage, we aim to determine these relationships, represented as links connecting image nodes, without determining the direction. The underlying principle is that as an image undergoes more modifications, its similarity to its original state decreases. This principle guides the calculation of dissimilarities across all image pairs to identify potential links.

Previous work has calculated image dissimilarity based on color distribution or by quantifying manufactured features like interest points. However, these image-matching methods, not originally developed for provenance tasks, often must adequately address the nuances of global or local manipulations. To address this problem, learning-based methods have been introduced to extract image features that accurately reflect the dissimilarities caused by various manipulations. However, existing provenance datasets typically need the specification of manipulated regions, making it challenging to train models directly at the image level to capture subtle manipulation traces. Current learning-based approaches often rely on manually created near-duplicate small patches with predefined manipulations. However, this approach can lead to sub-optimal performance due to potential misalignment with the types or intensity of manipulations in the original datasets. Moreover, it limits the scope to isolated patches, losing focus on adjacent regions or the entire image.

To overcome these limitations, we propose a novel method of learnable embedding that captures local and global transformations directly from the given data. This approach allows us to utilize near-duplicates directly from the provenance dataset while preserving the original topological proximity and other key underlying information. By doing so, we can better address the challenges of detecting localized and image-wide manipulations. The specifics of this method will be detailed in subsequent sections of this paper.

\subsubsection{Patch Embedding Extraction}
The learning-based method for link construction aims to encode transformation information into the extracted features properly. Intuitively, larger image dissimilarities should correspond to greater distances between extracted feature embeddings. Previous approaches, such as the one demonstrated by Li et al. \cite{li2021facial}, have proposed to encode images into an embedding space that minimizes the distance between closely related images while maximizing it for those with longer manipulation paths. As the image-level approach struggles to capture small but significant local changes, such as minor alterations from small donors, our work adopts a more reliable strategy: We split each image into patches and calculate the distances between corresponding patch embeddings. 

In our approach, all original images are uniformly rescaled to $224\times224$ pixels using bicubic interpolation techniques to standardize input dimensions and patch partition. The distance between any two images, $(I_{i},I_{j})$ is quantified by the pairwise patch distance as:
\begin{equation}
\delta(I_{i},I_{j})=\sum_{k}\delta(\mathbf{p}_{k}^{(i)},\mathbf{p}_{k}^{(j)}),
\label{eq:pairpatchdist}
\end{equation}
where the $d(\mathbf{p}_{k}^{(i)},\mathbf{p}_{k}^{(j)})$ denotes the Euclidean distance between corresponding $k$-th patch embeddings from $I_{i}$ and $I_{j}$ at the same spatial position: 
\begin{equation}
    \delta(\mathbf{p}_{k}^{(i)}, \mathbf{p}_{k}^{(j)})=\|\mathbf{p}_{k}^{(i)} - \mathbf{p}_{k}^{(j)}\|_2.
\end{equation}

\begin{figure}[t]
\centering
\includegraphics[width=\linewidth]{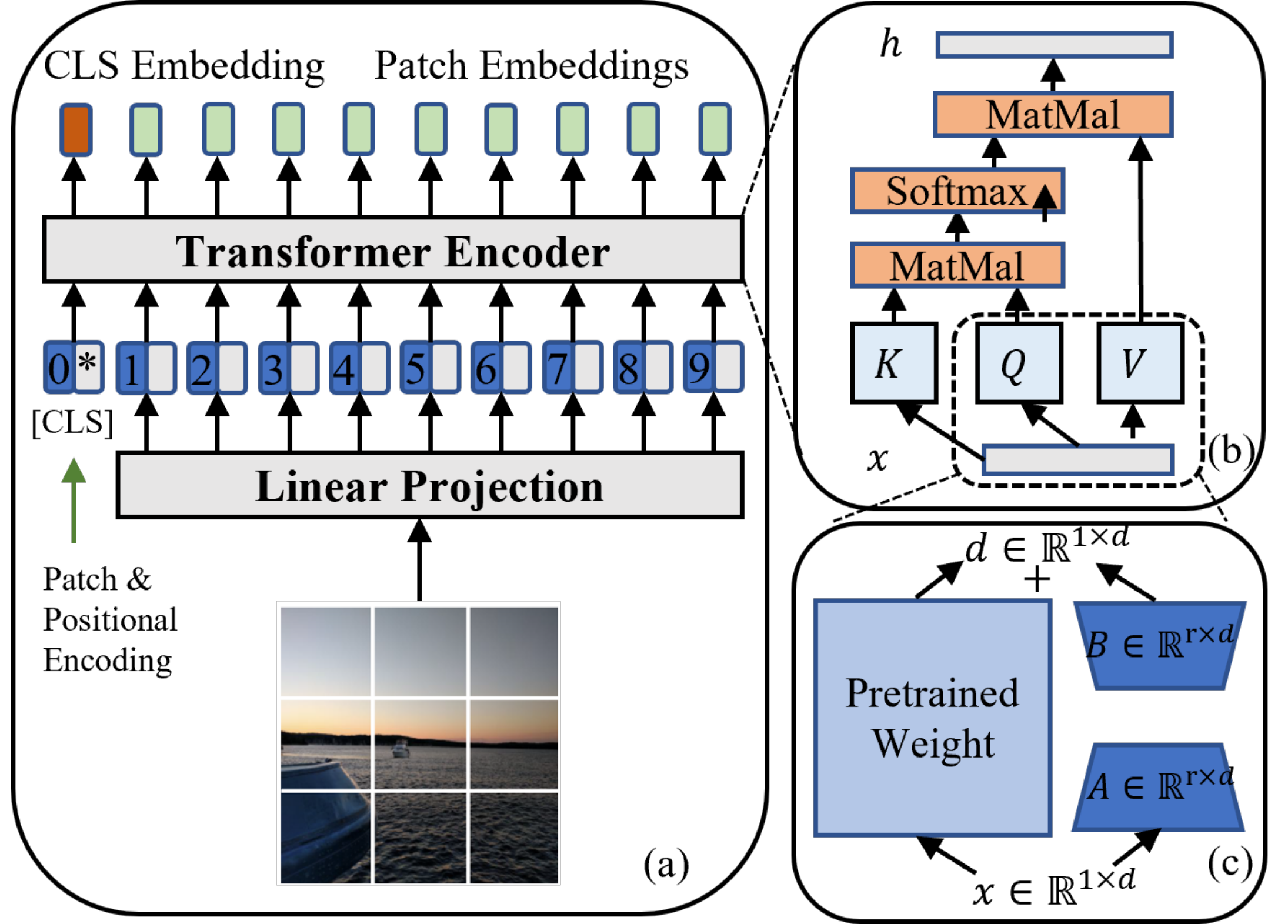}
\caption{Illustration of ViT-LoRA model. In the self-attention blocks of the encoder, the low-rank decomposition matrices are injected for fine-tuning, while the pre-trained weights of ViT are frozen. }
\label{ViTLoRA}
\end{figure}

To extract patch embeddings and preserve the inherent contextual understanding of images, we leverage the pre-trained ImageNet Vision Transformer (ViT) \cite{dosovitskiy2020image} as our backbone model, utilizing the patch embeddings from its last hidden state. As illustrated in Fig.\ref{ViTLoRA}, we incorporate Low-Rank Adaptation (LoRA) \cite{hu2021lora} layers into the attention blocks of the encoder for fine-tuning. This approach allows us to maintain parameter efficiency while making targeted adjustments to the output embeddings. Specifically, in each attention block, the output feature $h_{out}$ is derived from the input feature $h_{in}$ utilizing the original frozen ViT weights $W_{q/k/v}$ and two trainable low rank matrices:

\begin{equation}
    h_{out}=W_{q/k/v}h_{in} + BAh_{in},
\end{equation}
where $h_{in} \in \mathbb{R}^{1 \times d}$, $h_{out} \in \mathbb{R}^{d \times 1}$ and $B \in \mathbb{R}^{d \times r}$, $A \in \mathbb{R}^{r \times d}$ with $r << d$.
The ViT hidden state dimension $d$ is set to 768, and the rank $r$ is set to 16. As the positional encoded image patches pass through the transformer layers, the resulting output hidden states provide the patch embeddings used for calculating distances.

\subsubsection{Weighted Graph Distance Loss}
Manipulation techniques can be broadly categorized into two main classes: \textit{global transformations} and \textit{local transformations}. Global transformations affect the entire image, while local transformations influence specific regions. The path length between nodes in provenance graphs can reflect the overall dissimilarity between images. However, this path length cannot be directly applied to optimize patch embeddings due to the lack of information about the manipulated area. To address this challenge, we propose to leverage weights from a pre-trained model to guide the learning of patch embeddings. It enables the model to properly focus on the transformed patches, regardless of whether the manipulation is global or local. 

Specifically, while not explicitly designed to reflect distances in the provenance map, the pre-trained feature extractors still offer valuable insights into distinguishing manipulated regions. When an image undergoes global manipulation, all patch embeddings are transformed similarly in the feature space. Conversely, the transformation degree of patch embeddings in the affected areas will be greater for local manipulations than in unaltered regions.

During training, as Fig.\ref{Ovewview}(a) and Fig.\ref{Ovewview}(b) show, the learnable patch embeddings $\mathbf{p}$ of all filtered images are extracted along with the fixed patch embeddings $\bar{\mathbf{p}}$ from another frozen ViT module. For each image pair $(I_{i},I_{j})$, with $I_{i},I_{j} \in S$, we calculate the pairwise patch distances $\delta({\mathbf{p}_{k}^{(i)}}, {\mathbf{p}_{k}^{(j)}})$ between learned embeddings and $\delta({\bar{\mathbf{p}}_{k}^{(i)}}, {\bar{\mathbf{p}}_{k}^{(j)}})$ between fixed embeddings of patches at the same position. The weight of each patch pair is calculated by applying softmax to distance of pairwise fixed embeddings:
\begin{equation}
   w_{k}^{(ij)} = \frac{\text{exp}(\delta({\bar{\mathbf{p}}}_{k}^{(i)}, {\bar{\mathbf{p}}}_{k}^{(j)}))}
   {\sum_{t}\text{exp}(\delta({\bar{\mathbf{p}}}_{t}^{(i)}, {\bar{\mathbf{p}}}_{t}^{(j)}))}.
\end{equation}

\begin{figure}[t]
\centering
\includegraphics[width=\linewidth]{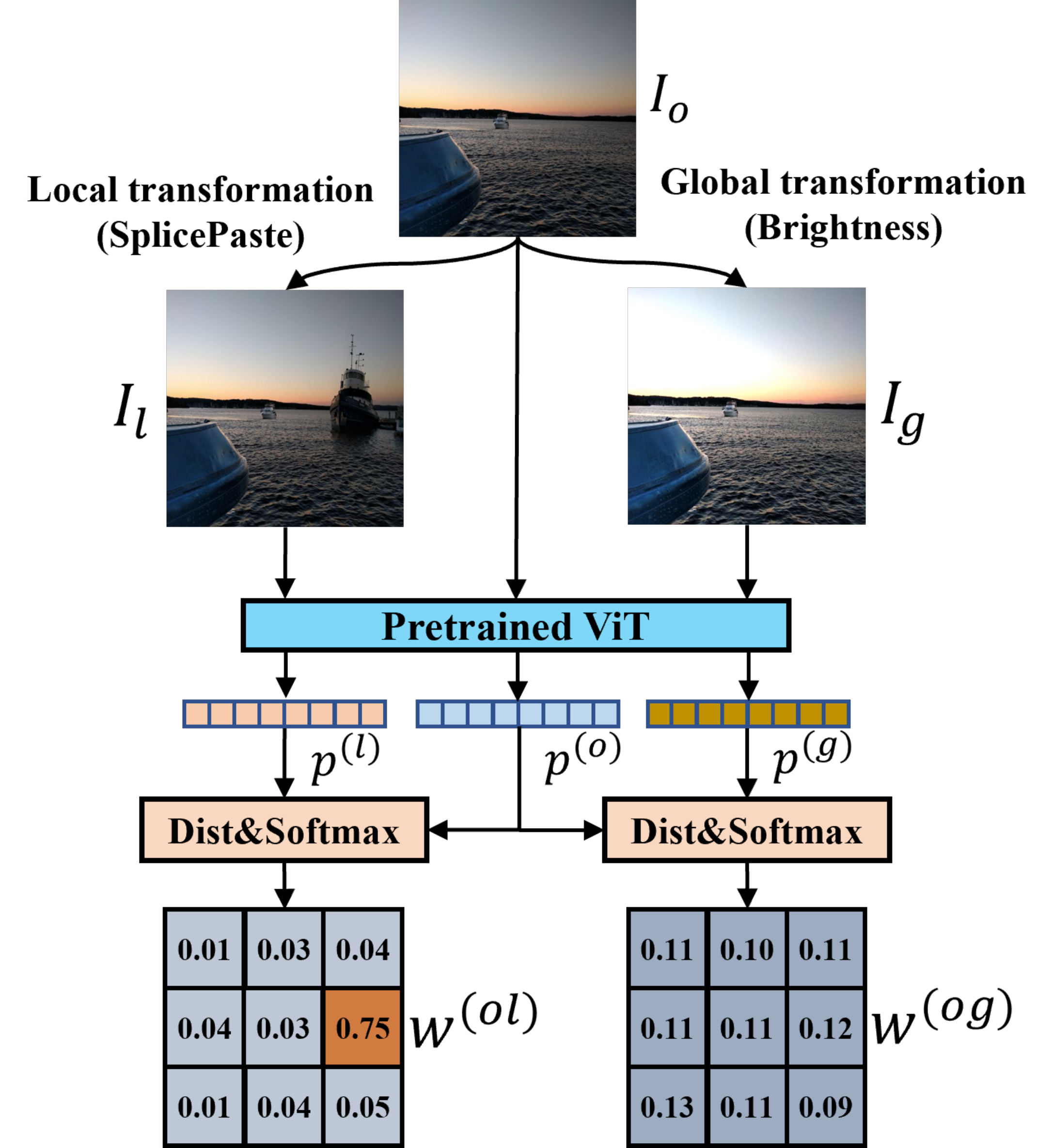}
\caption{Illustration of weights of patches from pre-trained ViT. For local transformation, the weights of all patches are similar. For local transformation, the weights of the manipulated regions are larger.}
\label{Weights of pairwise patches}
\end{figure}

As Fig.\ref{Weights of pairwise patches} depicts, the weights assigned to individual patches vary based on the region of the image modifications. For localized changes, the weights are significantly higher on the affected patches, enabling our model to focus on these areas of interest. In contrast, global transformations result in more uniform weights across all patches. By multiplying these weights with the corresponding pairwise patch distances, we compute a weighted distance for each image pair: 

\begin{equation}
   \delta_{w}(I_{i},I_{j})=\sum_{k}\delta(\mathbf{p}_{k}^{(i)},\mathbf{p}_{k}^{(j)})w_{k}^{(ij)}.
\end{equation}

From the perspective of each image, the one most similar to it typically has the highest probability of maintaining a link. For each image $I_i$, we apply normalization to all distances from other images to $I_i$. To facilitate optimization, we convert these normalized distances into probabilities of not having a link to the image: 
\begin{equation}
   \tilde{P}(I_{i},I_{j})= \frac{{\delta}_{w}(I_{i},I_{j})}{\sum_{l=0}^{n-1}{\delta}_{w}(I_{i},I_{l})},
\end{equation}
where $n$ is the number of images in each provenance graph.

To ensure that the learned embeddings accurately capture the dissimilarity and preserve the topological structure of the provenance graph in our embedding space, our objective is to align the ranking of probabilities with the distance of paths depicted in the ground truth data. Specifically, we aim to achieve a target where paths with fewer hops correlate with a lower probability value $\tilde{P}$. To achieve this, we minimize the following loss function:
\begin{equation}
   \mathcal{L}_u({PL},\tilde{P})=-\frac{\sum_{i=0}^{n-1}\sum_{m}\sum_{j=0}^{n-1}{PL}^m_i(j)\log(\tilde{P}(I_{i},I_{j}))}{n},
\end{equation}
${PL}$ is the ground path length, and $n$ is the number of images in the filtered image set. The ${PL}^m_i$ represents the $m$-th manipulation path that starts from node $i$, and ${PL}^m_i(j)$ is the number of steps between nodes $i$ and $j$ if a path exists from node $i$ to node $j$. Otherwise, it is set to $0$.

\subsubsection{Undirected Graph Construction}
After training, the patch embeddings extracted should be better suited for representing the transformation traces of images and effectively measuring the dissimilarity between them for predicting links. We use these embeddings to calculate distances between image pairs, constructing a dissimilarity matrix $D_u$ that encapsulates the relationships between all possible pairs of images from the filtered image set. Corresponding to Eq.\ref{eq:pairpatchdist}, the element $D_u(i,j)$ in the matrix equals to $d(I_{i},I_{j})$. Since this distance is unaffected by the order of the images—meaning $\delta(I_{i}, I_{j})$ = $\delta(I_{j}, I_{i})$, the resulting dissimilarity matrix $D_u$ is symmetric.

As previously discussed, the predicted provenance graph should be connected, with each node having at least one adjacent node with the lowest dissimilarity. To construct the undirected provenance graph based on this precondition, we apply Kruskal's Minimum Spanning Tree (MST) algorithm \cite{kruskal1956shortest} to the symmetric dissimilarity matrix $D_u$ to build the links. The algorithm operates by adding the edges between nodes in ascending order based on their dissimilarity values, ensuring no cycles are formed until all images are connected. The final structure of this graph is represented using an undirected adjacency matrix $\hat{A}_u$, where the element $\hat{A}_u(i,j)$ is set to 1 if there is a link between $I_i$ and $I_j$, and 0 otherwise.

\subsection{Direction Determination}
The direction of manipulation operations is another critical factor in provenance analysis. Several studies have attempted to determine it by comparing the mutual information of pixel values \cite{moreira2018image} or by utilizing forensic trace detectors \cite{zhang2020discovering}. However, recognizing the source and target of each transformation based solely on the visual content of image pairs presents significant challenges. This difficulty primarily arises from the fact that some transformations can be reversed, which renders the order ambiguous \cite{bharati2019beyond}. While other studies have explored the hidden relationships in metadata to reveal the order of manipulations \cite{bharati2019beyond}, the ease with which metadata can be manipulated and modified undermines the reliability of this method. 

We propose a more robust and accurate method beyond analyzing isolated image pairs. Specifically, we focus on learning from the overall structure of a provenance graph and perform direction determination using learnable precedence embeddings extracted from the graph structure masked attention module. The detailed implementations will be described in the subsequent subsections.

\subsubsection{Precedence Embeddings}
We leverage the asymmetry within the provenance graph to address the challenge of determining transformation directions. Recognizing that the provenance graph is a Directed Acyclic Graph (DAG), we exploit its topological ordering property. This property ensures that nodes can always be arranged in a descending sequence where each edge's start node precedes its end node. We propose assigning each image $I_{i}$ a distinct precedence embedding $\mathbf{e}^{(i)}$ in the latent space, representing its position within the provenance graph. For any image pair with a transformation relationship $(I_i \rightarrow I_j)$, the precedence embedding of $I_i$ should be closer to the most original image.

Given that the roots and leaves in the graph may be multiple, we incorporate the use of the virtual source and target nodes as illustrated in Fig.\ref{GraphMask}. The virtual source node represents the origin of all images, while the virtual target node represents the final manipulated version. In the descending sequence of the provenance graph, these virtual nodes serve as the initial starting point and final endpoint, respectively. Our model extracts the precedence embedding $\mathbf{e}$ for existing image nodes, and virtual embeddings $\mathbf{e}_{vs}$ and $\mathbf{e}_{vt}$ for virtual source and target nodes. Using these embeddings, we construct an auxiliary direction vector $(\mathbf{e}_{vs} - \mathbf{e}_{vt})$ in the latent embedding space. This vector provides a directional context that clearly defines forward movements in the graph, expressed as:
$(\mathbf{e}^i - \mathbf{e}^j)\cdot(\mathbf{e}_{vs} - \mathbf{e}_{vt})>0$.

\subsubsection{Graph Structure Masked Attention Module}
\begin{figure}[t]
\centering
\includegraphics[width=\linewidth]{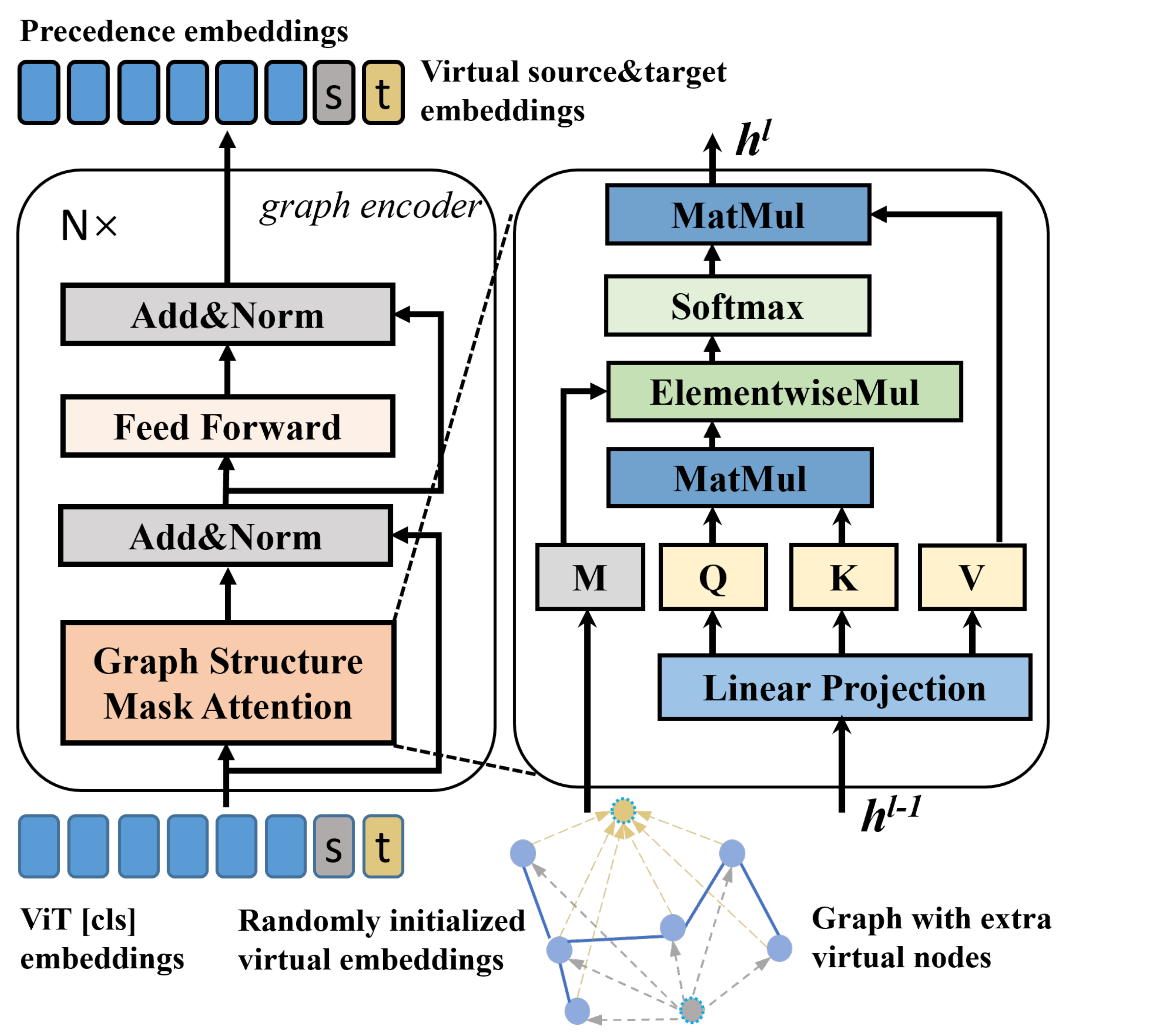}
\caption{Graph encoder with graph structure masked attention module. The designed module utilizes the structure of an undirected provenance graph to build the mask and control the scope of attention for each node. The inserted virtual source/target nodes are unidirectionally connected to all other nodes and assist in direction determination.}
\label{GraphMask}
\end{figure}

Beyond examining individual links or segments of the graph, we aim for the model to leverage the entire graph network to extract appropriate embeddings of image nodes to determine the direction of links. Numerous studies have explored using Graph Neural Networks (GNNs) \cite{hamilton2017representation, velivckovic2017graph,wu2020comprehensive} for information propagation across networks to generate node or edge embeddings. More recent efforts have adapted the structure of the Transformer \cite{yun2019graph,zhang2020graph,dwivedi2020generalization} to preserve proximity within graphs, demonstrating enhanced generalizability. In this work, as illustrated in Fig.\ref{Ovewview}(c), we employ a modified  Transformer encoder to convert image features into precedence embeddings. Specifically, as demonstrated in Fig.\ref{GraphMask}, we use the class embeddings derived from the second ViT-LoRA as the input, which condenses information across the entire image and is comprehensively aware of the provenance relationship. Additionally, we generate precedence and virtual node embeddings by diffusing image information through a modified attention mechanism, which will be discussed in detail. 

Unlike previous sequential data processing tasks like natural language processing, the order of input images is irrelevant to the underlying graph structure. It should not influence the resulting analysis in provenance tasks. Some works have utilized absolute graph encoding methods such as the Weisfeiler-Lehman algorithm \cite{zhang2020graph} and the use of Laplacian eigenvectors \cite{dwivedi2020generalization} in providing robustness against input permutations. Inspired by the concept of relative graph encoding implemented in GraphiT \cite{mialon2021graphit}, our approach moves away from traditional positional encoding within embeddings. Instead, we emphasize preserving topology information by integrating the graph structure directly into the attention mechanism. Given that the original self-attention mechanism interprets the input sequence as a fully connected network and applies attention to all input embeddings, we modify the attention blocks as Fig. ~\ref{GraphMask}:

\begin{equation}
   \text{GraphAttention}(Q,K,V,M)=(\text{softmax}(\frac{QK^T}{\sqrt{d_{out}}})\circ M)V,
\end{equation}
where $Q, K, V$ are query, key, and value matrix transformed from input embeddings by multiplying the corresponding weight matrices, $M$ is the graph structure mask to limit the scope of attention only on the connected nodes, $d_{out}$ is the output dimension which is the same as input in this work, and $\circ$ denotes the element-wise multiplication operation.


Before processing the image embeddings through the graph encoder, we append additional embeddings for the virtual nodes as illustrated in Fig.\ref{GraphMask}, which is randomly initialized. Since these virtual nodes connect to all other nodes, they need attention on all real nodes to facilitate the learning of aggregate representations of the provenance graph. It is crucial to note that to prevent undue information diffusion among the real image nodes via the virtual nodes, the virtual node attention implemented is unidirectional. With these considerations, we build the graph structure mask:
\begin{equation}
M = \alpha{\mathbf{I}} +\tilde{A}_u,
\end{equation}
\begin{equation}
\tilde{A}_u(i,j)
\begin{cases}
A_u(i,j) & \text{when  } i < n, \\
1 & \text{when  } n<=i<n+2 \text{  and  } i \neq j, \\
0 & \text{otherwise},
\end{cases}
\end{equation}
where the parameter $\alpha$ controls the ratio of attention a node directs towards itself versus its adjacent nodes and is set to $5$, $n$ is the number of filtered images, $\mathbf{I}$ is the identity matrix, $A_u$ is the adjacency matrix of an undirected graph. During the training phase, $A_u$ is constructed from the ground truth, while during inference, $A_u = \hat{A_u}$, which is derived from the link prediction process.  

\subsubsection{Directed Graph Construction}
After obtaining the precedence embeddings for each image node and the virtual source/target node embeddings, we utilize them to determine the direction of links. This is achieved by constructing a square direction matrix $D$ with dimensions matching the number of images in the provenance graph. The elements of $D$ are calculated as follows:
\begin{equation}
\hat{D}(i,j)= (\mathbf{e}^i - \mathbf{e}^j) \cdot (\mathbf{e}_{vs} - \mathbf{e}_{vt}).
\label{eq:hatD}
\end{equation}
Eq.\ref{eq:hatD} computes the difference in the projection of the source and target embeddings onto the auxiliary direction vector, which signifies the directionality of the transformation. Larger $D(i,j)$ suggests a higher probability that $I_i$ is the ancestor of $I_j$ in the transformation relationship.

During training, to ensure that the learned embeddings accurately reflect the position of the graph in the latent space, we minimize the following loss function:
\begin{equation}
\mathcal{L}_d(\hat{D},A_d) = \frac{\sum_{i=0}^{n-1}\sum_{j=0}^{n-1}\text{LeakyReLU}(\hat{D}(i,j)(A_d^\intercal-A_d))}{n^2},
\end{equation}
where $A_d$ is the directed adjacency matrix, with $A_d(i,j)=1$ indicating a directed link from node $i$ to node $j$. This loss function penalizes discrepancies between the predicted direction matrix $\hat{D}$ and the ground-truth represented by $A_d$.

With the predicted undirected adjacency matrix $\hat{A_u}$ from the link prediction stage and the direction matrix $\hat{D}$, the directed adjacency matrix $\hat{A}_d$ can be constructed through element-wise multiplication with the Heaviside step function $H$ applied to $\hat{D}$:
\begin{equation}
\hat{A}_d=H(\hat{D})\circ\hat{A_u}.
\end{equation}
The resulting matrix $\hat{A}_d$ represents the final directed provenance graph, where each entry $\hat{A}_d(i,j)=1$ indicates a directed connection from image $I_i$ to image $I_j$. 

To train the end-to-end provenance analysis model, we combine two loss functions corresponding to link prediction and direction determination stages into a single loss function and then minimize it to optimize both stages of the model simultaneously:
\begin{equation}
\mathcal{L} =\beta{\mathcal{L}_u} + \mathcal{L}_d,
\end{equation}
where $\beta$ is a coefficient used to balance the relative importance of these two sub-loss functions during the learning process, this integrated loss function ensures that the model effectively learns both to predict the connections between images and to accurately identify the source and target images of each link within the provenance graph.


\section{Experiments}

\subsection{Experimental setups}
\subsubsection{Implementation details}
We implemented the experiments on the Pytorch \cite{paszke2019pytorch} platform and trained concurrently for 30 iterations. The model utilizes a LeakyReLU loss function with a slope of 0.1 and the AdamW optimizer \cite{kingma2014adam} for training. The final loss function incorporates a balancing parameter $\beta = 0.1$. We maintain a constant learning rate of $1 \times 10^{-5}$ throughout the training period without decay. The model is trained on four RTX3080 GPUs with a batch size of 4. Our model's backbone includes a frozen ViT and two ViT-LoRA modules for image feature extraction. All components are initialized with weights from ViT-B pre-trained on the ImageNet-21k dataset. During the training phase, we derive graph masks used in the attention blocks directly from the ground-truth provenance graph. During inference, the graph mask is determined by the undirected adjacency matrix generated by the link prediction section of the model.

\subsubsection{Datasets}
The Open Media Forensic Challenge (OpenMFC) has released several provenance datasets \cite{guan2019mfc}, which include a large-scale collection of "world images" and "probe images." World images constitute a broad pool of ancestors or offspring to the probe or entirely irrelevant items. Probe images serve as the queries that guide the filtering process to identify and retrieve related images from the world dataset. To eliminate the influence of filtering, they provide all related images corresponding to the probe image for the oracle task. Each provenance graph may start from multiple root images, and the other images are manipulated from the roots with various image-editing tools. The datasets document the image modification history, ranging from local manipulations like copy-paste, removals, and donor changes to global manipulations such as compression, contrast modifications, scaling, sharpening, and blurring. 

Another provenance dataset is derived from Photoshop battles in the Reddit community. In these battles, community users start with one provided image and creatively create and modify different versions. This activity subsequently generates provenance graphs. Compared with OpenMFC datasets, Reddit dataset contains much more images in each provenance case, and most of the images are directly manipulated from one root image. Each datasets applied in the experiments provide ground-truth provenance graphs in nodes and directed links. We randomly select $70\%$ of cases for training and the remainder for evaluation. Due to partially disabled hyperlinks and missing images, our study only incorporates those cases that are still accessible and valid, ensuring that the provenance graph remains connected. The detailed information about the datasets is described below:

\begin{itemize}
    \item \textbf{NC2017-Dev1-Ver1} Dataset contains 394 provenance graphs and 6,498 world images. Each graph contains between 3 and 44 images. The number of edges per graph varies from 2 to 56, with an average of 19 edges per graph.
    \item \textbf{MFC18-Dev1-Ver2} Dataset contains 178 provenance graphs and 1,861 world images. Each graph contains between 5 and 42 images. The number of edges per graph varies from 5 to 46, with an average of 11 edges per graph.
    \item \textbf{Reddit} Dataset contains 118 provenance graphs and 3,092 world images. Each graph contains between 16 and 78 images. The number of edges per graph varies from 15 to 77, with an average of 48 edges per graph.
\end{itemize}

\subsubsection{Evaluation Metrics} 
In our study, we adopt the evaluation metrics specified by the OpenMFC to assess the accuracy of provenance graph construction. These metrics evaluate the consistency between the constructed provenance graph $G'$ and the ground-truth graph $G$. A higher metric score indicates a more accurate reconstruction of the original provenance history. We primarily employ two metrics: Vertex Overlap (VO) and Edge Overlap (EO). Both metrics calculate the F1 score, which is the harmonic mean of precision and recall, for the predicted nodes and edges, respectively:
\begin{equation}
    VO(G,G') = 2 \times \frac{|vt_{G'} \cap vt_{G}|}{|vt_{G'}|+|vt_{G}|}.
\end{equation}
\begin{equation}
    EO(G,G') = 2 \times \frac{|ed_{G'} \cap ed_{G}|}{|ed_{G'}|+|ed_{G}|}.
\end{equation}
Additionally, we use the Vertex Edge Overlap (VEO) to measure the overall graph overlap, providing a combined F1 score for both nodes and edges:
\begin{equation}
    VEO(G,G') = 2 \times \frac{|vt_{G'} \cap vt_{G}|+|ed_{G'} \cap ed_{G}|}{|vt_{G'}|+|vt_{G}|+|ed_{G'}|+|ed_{G}|}.
\end{equation}

The Vertex Overlap (VO) measures whether the method can correctly include all images in the graph, while the Edge Overlap (EO) assesses the accuracy of the connection between images. The Vertex Edge Overlap (VEO) combines both measures, providing an overall score of how well our predicted graph matches the ground-truth.  These metrics are applied to individual graphs; we then compute the average of these scores to obtain the final evaluation results.

\begin{table*}[t]
\centering
\caption{Provenance Graph Construction Result on NC2017-Dev1-Ver1 Dataset}
\begin{tabular}{l|ccc}
\hline
Solution      & VO   & EO   & VEO\\ \hline
SIFT\cite{zhang2020discovering}+MI\cite{moreira2018image} & $0.952 (\pm0.004)$  & $0.226 (\pm 0.015)$  & $0.580 (\pm 0.008)$      \\
SIFT\cite{zhang2020discovering}+Integrity\cite{zhang2020discovering} & $0.952 (\pm 0.004)$  & $0.231 (\pm 0.012)$  & $0.585   (\pm 0.006)$     \\
ResNet-50\cite{he2016deep}+Intergrity\cite{zhang2020discovering} & $1.000 (\pm 0.000)$ & $0.230 (\pm 0.021)$  & $0.610 (\pm 0.012)$    \\
ViT-B\cite{dosovitskiy2020image}+Intergrity\cite{zhang2020discovering} & $1.000 (\pm 0.000)$   & $0.233 (\pm 0.003)$  & $0.614 (\pm 0.002)$   \\
TAE\cite{bharati2021transformation}+Intergrity\cite{zhang2020discovering} & $1.000(\pm 0.000)$   & $0.250 (\pm 0.016)$   & $0.624 (\pm 0.010)$   \\
Ours & {$\textbf{1.000}(\pm 0.000)$}   & \textbf{$\textbf{0.289} (\pm0.015)$}   & \textbf{$\textbf{0.640} (\pm 0.008)$}   \\ \hline
\end{tabular}
\label{Table:NC2017}
\end{table*}

\begin{table*}[t]
\centering
\caption{Provenance Graph Construction Result on MFC18-Dev1-Ver2 Dataset}
\begin{tabular}{l|ccc}
\hline
Solution      & VO   & EO   & VEO\\ \hline
SIFT\cite{zhang2020discovering}+MI\cite{moreira2018image} & $0.942(\pm0.007)$  & $0.302(\pm0.004)$  & $0.619(\pm0.003)$      \\
SIFT\cite{zhang2020discovering}+Integrity\cite{zhang2020discovering}  & $0.942(\pm0.007)$ & $0.227(\pm0.032)$ & $0.617(\pm0.016)$     \\
ResNet-50\cite{he2016deep}+MI\cite{moreira2018image} & $1.000(\pm0.000)$ & $0.322(\pm0.018)$  & $0.668(\pm0.010)$      \\
ViT-B\cite{dosovitskiy2020image}+MI\cite{moreira2018image} & $1.000(\pm0.000)$   & $0.325(\pm0.022)$ & $0.669(\pm0.011)$    \\
TAE\cite{bharati2021transformation}+MI\cite{moreira2018image} & $1.000(\pm0.000)$   & $0.347(\pm0.019)$   & $0.680(\pm0.010)$    \\
Ours & $\textbf{1.000}(\pm0.000)$   & $\textbf{0.424}(\pm0.013)$  & $\textbf{0.717}(\pm0.007)$   \\ \hline

\end{tabular}
\label{Table:MFC18}
\end{table*}

\begin{table*}[t]
\centering
\caption{Provenance Graph Construction Result on Reddit Dataset}
\begin{tabular}{l|ccc}
\hline
Solution      & VO   & EO   & VEO\\ \hline
SIFT\cite{zhang2020discovering}+MI\cite{moreira2018image} & $0.967(\pm0.005)$  & $0.106(\pm0.006)$  & $0.542(\pm0.004)$      \\
SIFT\cite{zhang2020discovering}+Integrity\cite{zhang2020discovering}  & $0.967(\pm0.005)$ & $0.097(\pm0.017)$ & $0.537(\pm0.014)$     \\
ResNet-50\cite{he2016deep}+MI\cite{moreira2018image} & $1.000(\pm0.000)$ & $0.082(\pm0.010)$  & $0.550(\pm0.008)$      \\
ViT-B\cite{dosovitskiy2020image}+MI\cite{moreira2018image} & $1.000(\pm0.000)$   & $0.078(\pm0.013)$ & $0.551(\pm0.012)$    \\
TAE\cite{bharati2021transformation}+MI\cite{moreira2018image} & $1.000(\pm0.000)$   & $0.076(\pm0.017)$   & $0.548(\pm0.013)$    \\
Ours & $\textbf{1.000}(\pm0.000)$   & $\textbf{0.139}(\pm0.019)$   & $\textbf{0.577}(\pm0.015)$  \\ \hline
\end{tabular}
\label{Table:Reddit}
\end{table*}

\subsection{Data Augmentation}
In our research on image provenance analysis, we rely upon specialized data augmentation techniques to enhance the diversity and volume of our training datasets. This process differs from traditional computer vision tasks, requiring consideration of both images and the underlying graph topology representing correct relationships. The designed augmentation strategy should preserve the integrity of the original provenance information within graph structures, and avoid modifications that could obscure original provenance paths or detrimentally impact model performance.

The data augmentation methodology encompasses two approaches. First, we modify existing images and introduce additional branches from specific nodes within the provenance graphs. This technique increases graph complexity and extends manipulation chains, creating more intricate structures. We leverage global manipulation techniques: brightness, saturation, contrast, sharpness, and blur, which are outlined in the the OpenMFC datasets, to introduce new nodes, effectively. Second, we selectively reduce graph size by removing nodes with a single ancestor and their associated links, directly connecting the ancestor node to any offspring of the removed node. This method generates smaller graphs with links representing multiple modification steps, improving our framework's performance under varied structural conditions. Throughout this process, we provide a broader spectrum of example cases for enhancing the robustness and effectiveness of our provenance analysis framework with preservation of original provenance information, ensuring that our augmented datasets contribute meaningfully to the model's learning without introducing misleading or distorted relationships.

\subsection{Baseline Approaches}
To assess the effectiveness of the proposed framework, we conduct a comprehensive comparison with other available vision-content-based methods. In the prior art, the evaluation process considers the two-stage of provenance graph construction: link prediction and direction determination. For each stage, we select several approaches as evaluation baselines.

In link prediction for image provenance, baseline techniques are categorized into two main types: interest point detectors and learned feature descriptors. Interest point detectors, such as SIFT and SURF, are employed in research by Moreira \cite{moreira2018image}, and Zhang \cite{zhang2020discovering} to identify distinctive image features. Following the algorithm in \cite{zhang2020discovering}, the dissimilarity between image pairs is assessed by counting the top 2000 SIFT keypoints matches.

Learned feature descriptors, conversely, encapsulate image content into embeddings using machine learning techniques. For our comparative analysis, we select two image feature extraction models pre-trained on ImageNet \cite{deng2009imagenet}: ResNet-50 \cite{he2016deep} and the base Vision Transformer (ViT-B) \cite{dosovitskiy2020image}. Additionally, we assess the Transformation-Aware Embeddings (TAE) framework \cite{bharati2021transformation}, specifically designed to discern transformations between images. TAE employs a four-way Siamese neural network trained with quadruplet image samples and the rank-based Edit Sequence Loss. Due to the unavailability of their handcrafted dataset, we extract quadruplet samples directly from corresponding training cases. To ensure evaluation consistency, we adhere to the settings described by Bharati et al. \cite{bharati2021transformation} for both ResNet-50 and TAE frameworks. The ViT-B framework generates patch embeddings in the same manner as our proposed method. The dissimilarity between images is determined by summing the pairwise distances of patch embeddings for all these approaches. 

We then examine two established methods to determine the direction of manipulation between pairs of images. The first approach, rooted in information theory \cite{moreira2018image}, calculates the mutual information (MI) between image pairs. This method aligns matched interest points from one image to another, mapping related regions from image $I_i$ onto $I_j$. The method detects asymmetries indicative of an ancestor-offspring relationship by assessing the mutual information within pixel values in aligned regions. It leverages variations in forward and backward homography to identify potential donor images. 

Additionally, we explore another baseline method in \cite{zhang2020discovering} that employs a trained forensic tool to evaluate image integrity. This approach assigns an integrity score to each image, operating under the assumption that manipulated images typically exhibit lower scores compared to their original versions. The manipulation direction is then determined by comparing these integrity scores among linked images. For our experiment, we utilize the state-of-the-art framework proposed by Guillaro \cite{guillaro2023trufor}, employing its pre-trained integrity scorer to conduct evaluations and draw comparisons.

\section{Experimental Results}

\subsection{Intra-Dataset Evaluation}
The evaluation of the proposed image provenance analysis framework yields insights into its performance relative to established baseline methods. The assessment encompasses both stages of provenance graph construction and the reconstructed provenance graph is represented as a directed adjacency matrix compared to the ground-truth for all methods. To ensure a thorough assessment, we conduct experiments on the following challenging datasets: NC2017-Dev1-Ver1, MFC18-Dev1-Ver2, and Reddit. We present evaluation results using the metrics Vertex Overlap (VO), Edge Overlap (EO), and Vertex-Edge Overlap (VEO) to show the accuracy of predicted graphs. For baseline approaches that focus solely on link prediction and constructing undirected graphs, we pair them with the best previous direction determination methods to generate comparable directed graphs. The following tables present the results of these best combinations alongside our framework, providing average scores and standard deviations from three evaluation runs, with the best results highlighted in bold.

Table \ref{Table:NC2017} and \ref{Table:MFC18} show the provenance graph construction results on the NC2017-Dev1-Ver1 and MFC18-Dev1-Ver2 datasets. Our proposed framework outperforms all other methods across all metrics. Notably, SIFT-based methods fail to retrieve all images for graph building, resulting in Vertex Overlap (VO) values below 1. This limitation stems from insufficient matched interest points between image pairs, a critical defect in interest point-based methods. In contrast, all learning-based feature extractors, which utilize patch embeddings to form the dissimilarity matrix, successfully incorporate all nodes into the graphs, achieving perfect VO scores.

The Edge Overlap (EO) metric is more crucial in differentiating framework performance. Our method achieves the best EO result, showing a relative improvement of $15.6\%$ on NC2017-Dev1-Ver1 and $22.2\%$ on MFC18-Dev1-Ver2 compared to other baseline approaches. This enhancement demonstrates our approach's superior understanding of various manipulation types and transformation directions.

Table \ref{Table:Reddit} shows the evaluation results on the Reddit dataset. We observe a performance decline across all methods compared to other datasets. This degradation is attributed to the larger number of images and edges in each provenance graph, which inherently complicates the task of accurate link prediction. Despite these challenges, proposed method shows robust performance, achieving a $31.1\%$ improvement over the best baseline approach. This observation underscores the effectiveness of our framework in learning transformation relationships from graphs, enabling it to handle more images with greater accuracy. Our method's ability to maintain superior performance on this challenging dataset also highlights its robustness in handling more complicated structures.

\begin{figure}[t]
\centering
\begin{subfigure}[b]{0.48\textwidth}  
    \centering
    \includegraphics[width=\textwidth]{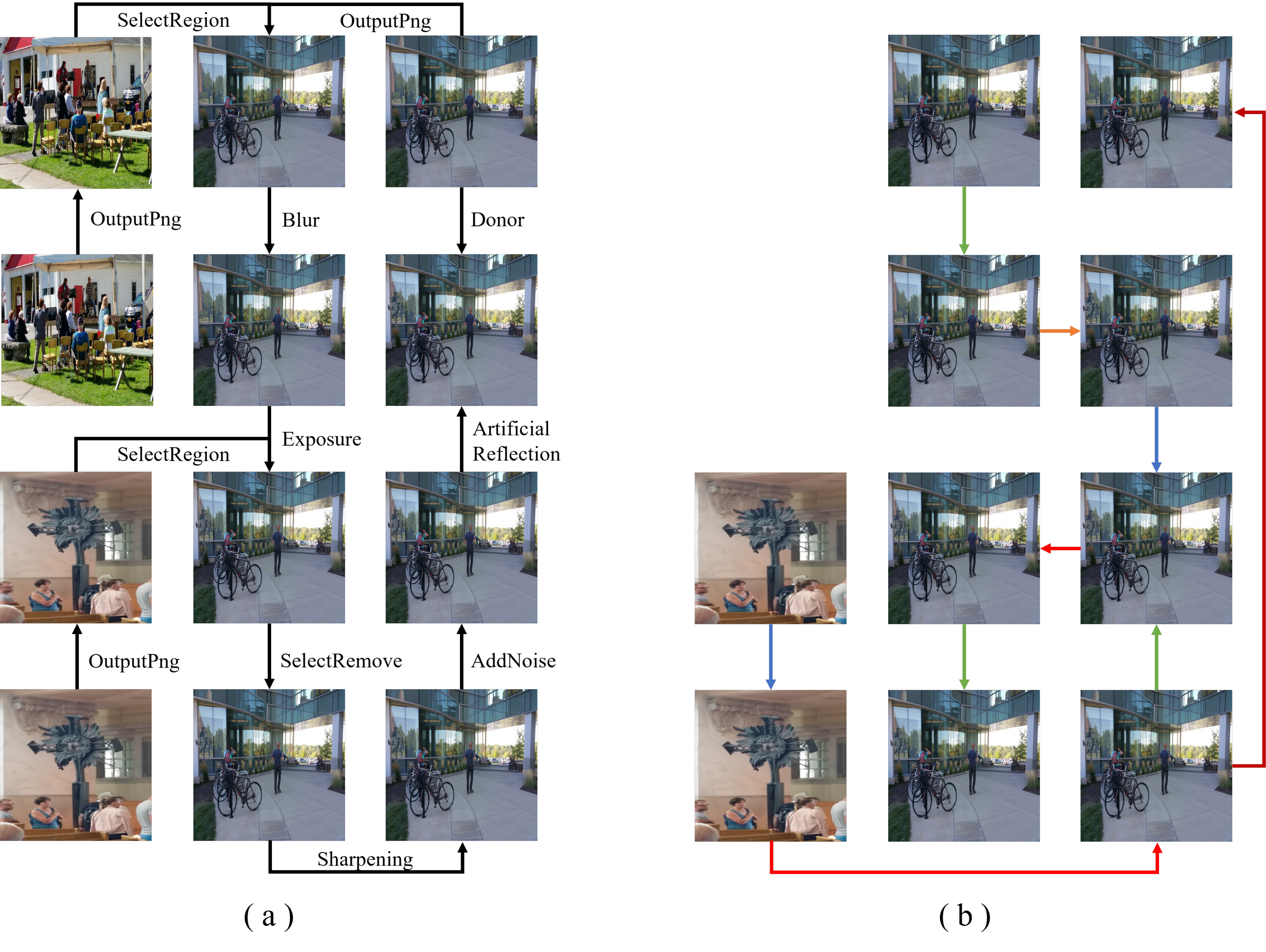}
    \label{fig:sub1}
    
\end{subfigure}

\begin{subfigure}[b]{0.48\textwidth}
    \centering
    \includegraphics[width=\textwidth]{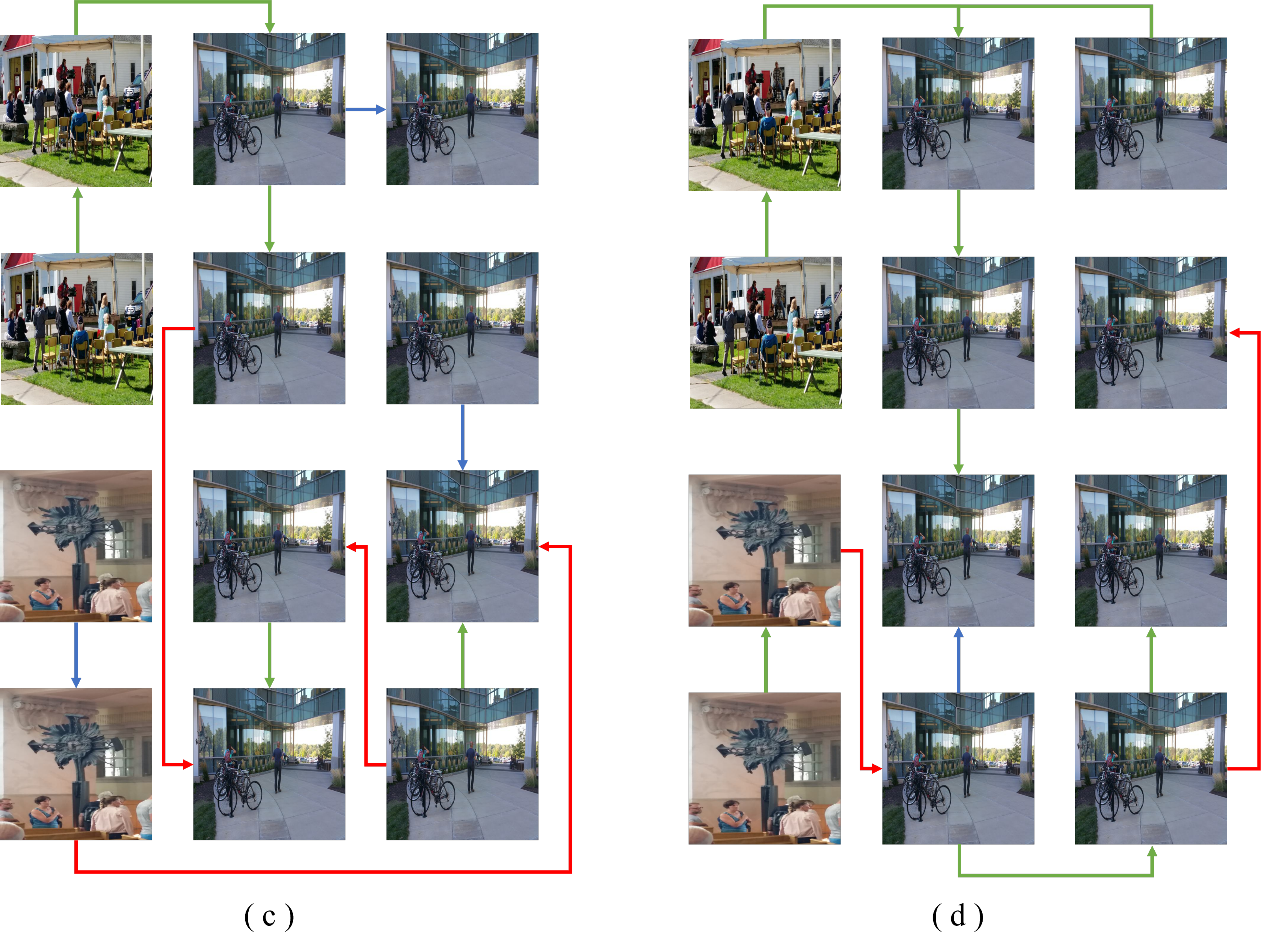}
    \label{fig:sub2}
\end{subfigure}
\caption{Qualitative results for directed provenance graphs. (a) shows the ground-truth graph. (b) the baseline result graph uses SIFT for link detection and integrity scorers for direction determination. (c) is the baseline result generated by TAE and integrity scorers. (d) depicts the result of our framework. Link colors indicate prediction accuracy: green represents correct connections and directions; blue shows correct connections with incorrect directions; red denotes non-existent connections in the ground truth.}
\label{fig:Qualitative Result MFC18}
\end{figure}

To better understand our framework's performance, we present a qualitative analysis of a provenance case in Fig.\ref{fig:Qualitative Result MFC18}. This visual comparison illustrates the effectiveness of our approach relative to baseline methods. Fig.~\ref{fig:Qualitative Result MFC18} (a) displays the ground-truth provenance graph, with arrows indicating transformation directions and labels specifying manipulation types between image pairs. Fig.~\ref{fig:Qualitative Result MFC18} (b) and (c) show results from baseline methods: SIFT and TAE with integrity scorers for direction determination, respectively. Fig.~\ref{fig:Qualitative Result MFC18} (d) presents the result of our proposed framework.  Unlike the SIFT-based method, which omits some images, our approach successfully incorporates all images into the graph structure. Moreover, our method achieves higher accuracy in connection establishment and direction determination than baseline approaches.

\begin{figure}[t]
\centering
\includegraphics[width=\linewidth]{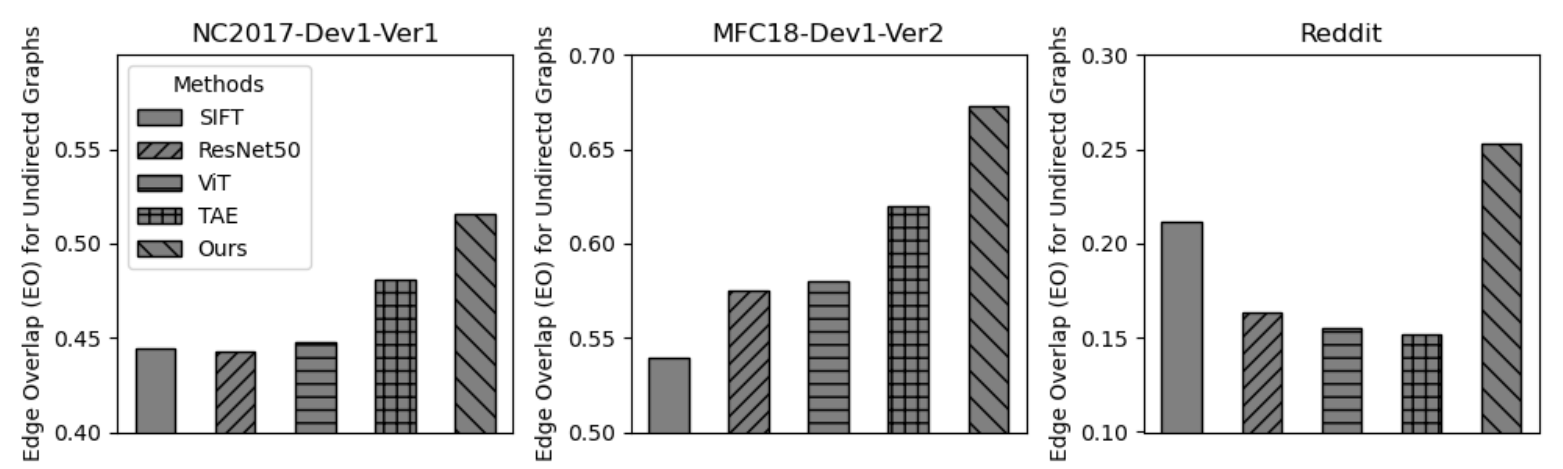}
\caption{Ablation experiments for the link prediction stage.}
\label{Ablation:Link}
\end{figure}

\begin{figure}[t]
\centering
\includegraphics[width=\linewidth]{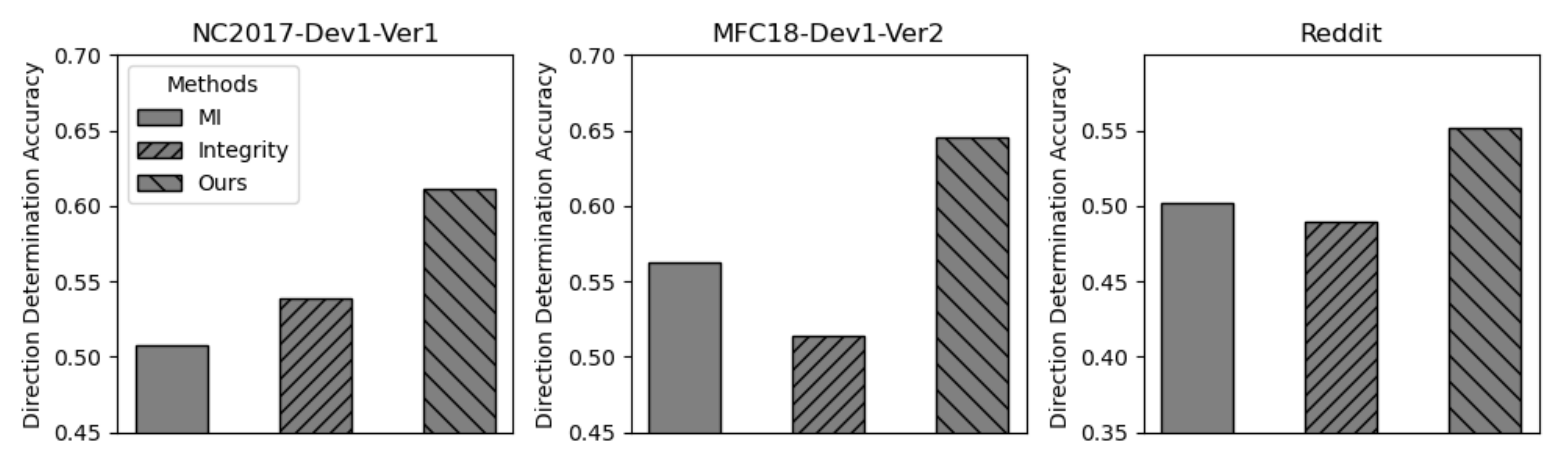}
\caption{Ablation experiments for the direction determination stage.}
\label{Ablation:Direction}
\end{figure}

\subsection{Cross-Dataset Evaluation}
Due to the uncontrollable environments in real-world application scenarios, the trained model inevitably encounters unseen domain data when deployed, resulting in poor testing performance. In the context of image provenance analysis, cross-dataset evaluation presents a persistent challenge due to the variability in manipulation techniques and graph topologies across different datasets. In our cross-dataset evaluation experiments, we train both TAE and our method on the NC2017-Dev1-Ver1 dataset while maintaining the pre-trained weights of ResNet50 as the benchmark. 

We test the trained models on the other two unseen datasets as illustrated in Table \ref{Table:cross-mfc18} and \ref{Table:cross-reddit}. Our framework, which relies on graph structure for prediction, experiences certain performance degradation compared to intra-dataset evaluations. This degradation is expected and can be attributed to the differences in image manipulation techniques, graph complexities, and data distributions between the training and testing datasets. Despite this challenge, our framework outperforms previous arts, highlighting the outstanding generalizability of our approach even when faced with unforeseen data structures and manipulation techniques.

The superior cross-dataset performance of our approach can be attributed to several factors. Firstly, our local-global joint learning scheme allows the model to capture both fine-grained image features and broader structural patterns, making it more robust to variations in manipulation techniques. Secondly, the graph structure masked attention module enables the framework to adapt to different graph topology, enhancing its performance across datasets with varying complexities. These results highlight the outstanding generalizability of our approach even when faced with unforeseen data structures and manipulation techniques. This robustness is particularly valuable in real-world applications where the nature of image manipulations and provenance graph structures may be unpredictable.

\begin{table}[t]
\centering
\caption{Cross-Dataset Evaluation Result on MFC18-Dev1-Ver2 Dataset}
\begin{tabular}{l|cc}
\hline
Solution      & EO   & VEO\\ \hline
ResNet-50\cite{he2016deep}+MI\cite{moreira2018image}  & $0.325(\pm0.002)$  & $0.669(\pm0.001)$      \\
TAE\cite{bharati2021transformation}+MI\cite{moreira2018image}& $0.337(\pm0.004)$   & $0.674(\pm0.003)$    \\
Ours & $\textbf{0.356}(\pm0.015)$   & $\textbf{0.683}(\pm0.007)$  \\ \hline
\end{tabular}
\label{Table:cross-mfc18}
\end{table}

\begin{table}[t]
\centering
\caption{Cross-Dataset Evaluation Result on Reddit Dataset}
\begin{tabular}{l|cc}
\hline
Solution      & EO   & VEO\\ \hline
ResNet-50\cite{he2016deep}+MI\cite{moreira2018image}  & $0.079(\pm0.001)$  & $0.553(\pm0.001)$      \\
TAE\cite{bharati2021transformation}+MI\cite{moreira2018image}& $0.087(\pm0.012)$   & $0.558(\pm0.008)$    \\
Ours & $\textbf{0.104}(\pm0.004)$   & $\textbf{0.565}(\pm0.002)$  \\ \hline
\end{tabular}
\label{Table:cross-reddit}
\end{table}

\subsection{Ablation Study}
We conduct ablation experiments to study the impacts of modules in each stage of provenance graph construction. For the link prediction stage, we compared the results of our sub-framework with four baseline approaches using the edge overlap metric without considering the direction. Fig.~\ref{Ablation:Link} illustrates the results of these ablation experiments across three datasets. Our analysis reveals that previous learning frameworks perform well on OpenMFC datasets but are less competitive with interest point methods on the Reddit dataset, which contains lots of versions with donors and entirely different backgrounds. Notably, our framework demonstrates significant improvements compared to other approaches across all datasets, underscoring its efficiency in predicting correct transformation relationships. The superior performance of our method can be attributed to its ability to capture both local and global features, allowing it to identify subtle transformations that might be missed by other approaches. 

Besides, with given ground-truth undirected graphs, we perform ablation experiments for the direction determination stage and analyze the accuracy of predicted directions. As illustrated in Fig.~\ref{Ablation:Direction}, our framework achieved the best performance in accurately predicting the direction. However, the improvement on the Reddit dataset is not as pronounced as on the other two datasets due to the distinctiveness of its graph topology, which is nearly star-shaped with numerous branches. This suggests that while our method excels in most scenarios, there is still room for improvement in some structures. Nevertheless, our framework still demonstrates superior performance, which can be attributed to the designed graph structure masked attention module. This module enables our framework to capture both fine-grained image relationships and broader structural patterns in the provenance graphs, allowing for effective analysis in complex topologies. In conclusion, the effectiveness of our approach in both stages underscores the importance of considering both image content and graph structure in provenance analysis tasks.

\begin{figure}[t]
\centering
\includegraphics[width=\linewidth]{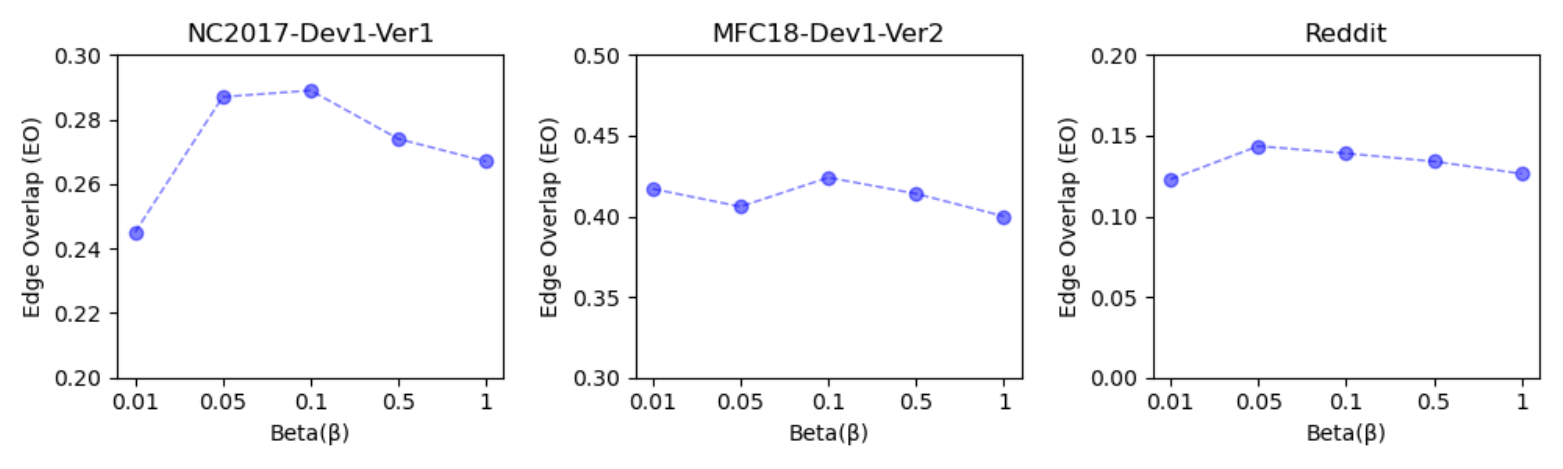}
\caption{Results of ablation experiments for coefficient in loss function.}
\label{Ablation:Coefficient}
\end{figure}

Furthermore, we examine the impact of varying the coefficient $\beta$ in our loss function. We test various values to determine the optimal balance between the components of the loss function. The findings, depicted in Fig.~\ref{Ablation:Coefficient}, reveal that $\beta=0.1$ provides the best results on the OpenMFC dataset, indicating a good balance between loss components, enhancing reliability and performance on both stages. However, the Reddit dataset shows slightly better performance at $\beta=0.05$, reflecting its specific requirements for handling more diverse manipulations and complex visual content variations.

\begin{table}[t]
\centering
\caption{Ablation Study Results for Data Augmentation on All Datasets in Link Prediction Stage}
\begin{tabular}{l|cc}
\hline
Dataset      & Original  & Augmented\\ \hline
NC2017-Dev1-Ver1 & $0.504(\pm0.011)$   & $0.516(\pm0.008)$  \\ 
MFC18-Dev1-Ver2 & $0.662(\pm0.015)$   & $0.673(\pm0.021)$  \\ 
Reddit & $0.251(\pm0.007)$   & $0.253(\pm0.010)$   \\ \hline
\end{tabular}
\label{Table:ablation-aug-link}
\end{table}

\begin{table}[t]
\centering
\caption{Ablation Study Results for Data Augmentation on All Datasets in Direction Determination Stage}
\begin{tabular}{l|cc}
\hline
Dataset      & Original  & Augmented\\ \hline
NC2017-Dev1-Ver1 & $0.586(\pm0.007)$   & $0.612(\pm0.013)$  \\ 
MFC18-Dev1-Ver2 & $0.621(\pm0.022)$   & $0.646(\pm0.009)$  \\ 
Reddit & $0.544(\pm0.007)$   & $0.552(\pm0.017)$   \\ \hline
\end{tabular}
\label{Table:ablation-aug-direction}
\end{table}

To evaluate the impact of our data augmentation techniques on the performance of the image provenance analysis framework, we conducted experiments on both the original datasets and those augmented through the specialized two-stage methods, as previously discussed. As detailed in Table \ref{Table:ablation-aug-link}, the model trained on augmented data exhibited higher performance in the link prediction stage using the metric of undirected edge overlap. This suggests that the additional branch images with various random global manipulations enhance the model's ability to capture transformation information and identify potential links between nodes. The performance increment for the Reddit dataset was less pronounced, possibly due to its already adequate image count and the presence of manipulation types not covered by our augmentation methods.

We also evaluated the accuracy of direction determination, where the effects of data augmentation were notably positive, as shown in Table \ref{Table:ablation-aug-direction}. By introducing diverse graph structures through our augmentation techniques, we enabled the model to learn from a broader range of scenarios. This diversity in training data proved especially beneficial for handling the intricate task of determining the directional flow of image transformations. The improved performance suggests that our augmentation strategy effectively enhances the model's ability to generalize across various structural conditions, thereby significantly boosting its predictive accuracy and robustness in discerning both the relationship and the directionality of image manipulations within provenance graphs.

\section{Conclusions and Future Work}
Unveiling the intent behind the spread of manipulated images on social media presents a significant challenge. Image provenance analysis offers an effective solution by tracing the manipulation history through visual graphs, thereby revealing connections among related images. This paper introduces an end-to-end framework designed to analyze the content of individual images and leverage the structural information within these provenance graphs for high-accuracy predictions of directed provenance relationships.

Our approach calculates the dissimilarity between images in the link prediction phase by analyzing the distance between pairwise patch embeddings. To enhance the graph construction performance, we incorporate weights from a pre-trained model into the patches to highlight manipulated regions and utilize transformation paths for optimization. During the direction determination phase, we integrate the graph's topology with an attention mask mechanism and introduce virtual nodes to improve the prediction of potential manipulation directions. 

In conclusion, our proposed learning scheme that encodes graph structures into vision transformers has significantly improved image provenance graph construction performance. Extensive ablation experiments have demonstrated the effectiveness of the designed components, used hyper-parameters, and adopted data augmentation strategy. The cross-dataset evaluations demonstrate that existing methods still suffer from limited generalizability, primarily due to significant deficiencies in training data. While our proposed method has mitigated the problem to some extent, enhancing the model's generalization capability remains a grand challenge. 

In future work, we aim to develop a more balanced and comprehensive dataset that encompasses a wider variety of manipulation techniques, including advanced deepfake technologies. Additionally, we plan to design a new framework that identifies the transformation types and enhances the richness of information within the provenance graph. We believe these future research goals will better fit real-world application scenarios and significantly improve the accuracy and generalizability of image provenance analysis.


\ifCLASSOPTIONcaptionsoff
  \newpage
\fi

\bibliographystyle{IEEEtran}
\bibliography{refs}

\end{document}